\documentclass[12pt]{article}
\usepackage{amsfonts,amsmath,amssymb,epsf}
\usepackage{graphicx,color}
\newcommand{\be}{\begin{equation}}
\newcommand{\ba}{\begin{eqnarray}}
\newcommand{\ea}{\end{eqnarray}}
\newcommand{\ee}{\end{equation}}

\newcommand{\no}{\nonumber \\}

\begin{document}

\begin{titlepage}

\begin{flushright}
\begin{tabular}{l}
KUNS-2269 \\
OU-HET-669-2010
\end{tabular}
\end{flushright}
\vspace*{3cm}
    \begin{Large}
    \begin{bf}
       \begin{center}
         {Geometric entropy and third order phase transition in $d=4$ $\mathcal{N}=2$ SYM with flavor} 
       \end{center}
    \end{bf}   
    \end{Large}
\vspace{1cm}

   \begin{large}
\begin{center}
{Mitsutoshi Fujita}$^a$\footnote    
{e-mail address : mfujita@gauge.scphys.kyoto-u.ac.jp} and
{Hiroshi Ohki}$^b$\footnote{e-mail address : ohki@het.phys.sci.osaka-u.ac.jp}

\end{center}
\end{large}

      \vspace{1cm}

\begin{center}\it{$^a$Department of Physics, Kyoto University, Kyoto 606-8502, JAPAN}\\ 
\it{$^b$Department of Physics, Graduate School of Science Osaka University, Toyonaka, Osaka 560-0043, Japan}\\

\end{center}

\vspace{1cm}

\begin{abstract}
We analyze the phase structure of $d=4$ $\mathcal{N}=2$ large $N$ SYM theory with flavor on $S^1\times S^3$ by using geometric entropy as an order parameter.  We introduce chemical potential conjugate to global $U(1)$ symmetry and find the third order phase transition at finite density by using the geometric entropy as the order parameter. We also find that the geometric entropy at the finite density has an interesting behavior at low temperature and for large $N_f$.
\end{abstract}
\vfill 

\end{titlepage}
\vfil\eject

\setcounter{footnote}{0}

\section{Introduction}
Geometric entropy~\cite{taka} was defined as the von-Neumann (information) entropy with Hamiltonian associated with the space translation. The definition of this geometric entropy was slightly different from that of the entanglement entropy~\cite{HLW,CC,Area} which appeared in the condensed matter physics: geometric entropy was related to the entanglement entropy by double Wick rotation. However, geometric entropy becomes more convenient observable to investigate finite temperature systems including the strongly coupled Quark Gluon Plasma (QGP) and the weakly coupled deconfined phase of Yang-Mills theories. And particular interest is to investigate the phase structure of large $N$ QCD at a finite density by using geometric entropy as an order parameter. 

 In the context of the gauge/gravity correspondence~\cite{ma,gkp,wi},
geometric entropy was introduced as a new order parameter to analyze
confinement/deconfinement transition of $d=4$ large $N$ $SU(N)$
Yang-Mills theory on $S^3$ at finite temperature~\cite{taka}. In the
gauge theory side, free Yang-Mills theory~\cite{sundberg,aharony,Aha2}
to mimic QCD has a natural scale, the radius $R$ of $S^3$, which is
small enough to believe asymptotic freedom:   $R\Lambda_{QCD}\ll 1$. And the confinement/deconfinement phase transition takes place if $R\Lambda_{QCD}$ is of order $1$.

In the dual gravity side, geometric entropy of the dual strongly coupled Yang-Mills theory\footnote{Interestingly, lattice theory results and the soft-wall AdS/QCD results~\cite{Kar2}-\cite{Fujjj2} show that QCD is more like strong coupling. } obeys the area law formula. Namely, it is proportional to a minimal surface stretching on the $AdS$ background. The first order confinement/deconfinement phase transition could also be seen holographically by using geometric entropy as an order parameter and the phase structure was consistent with the gauge theory results (see also the holographic calculation \cite{RT}-\cite{Nis} of the entanglement entropy based on the AdS/CFT correspondence).

In this paper, we analyze the phase structure 
of free $\mathcal{N}=2$ large $N$ Super Yang-Mills (SYM) theory with flavor on $S^1\times S^3$  by using geometric entropy as an order parameter. It is known that free $\mathcal{N}=2$ SYM with flavor can approximately be reduced to the Gross-Witten type matrix model~\cite{hliu}, which can describe the third-order phase transition~\cite{gross}-\cite{Han} instead of the first order phase transition of the large $N$ gauge theory. We also introduce chemical potential conjugate to global $U(1)$ symmetry and investigate the phase structure at a finite density. It is interesting to consider the application to the gauge/gravity correspondence~\cite{Karc}-\cite{Mate}. However, $\mathcal{N}=2$ SYM theory is in general not conformal and so the application seems not to be straightforward except the conformal case $N_f=2N$ or  $N_f\ll N$.
 
The content of this paper is as follows.
In section 2, we have shortly the review of geometric entropy. In section 3, we briefly review the partition function of free $\mathcal{N}=2$ SYM theory with flavor on $S^1\times S^3$. And we discuss the phase structure of this theory by obtaining free energy and Polyakov loop vev. In section 4, we capture the third order phase transition of $\mathcal{N}=2$ SYM theory with flavor by using geometric entropy. In section 5, we introduce chemical potential conjugate to two global $U(1)$ groups and generalize our model to the finite density system.
\section{Geometric entropy}

\begin{figure}[htbp]
   \begin{center}
     \includegraphics[height=4cm]{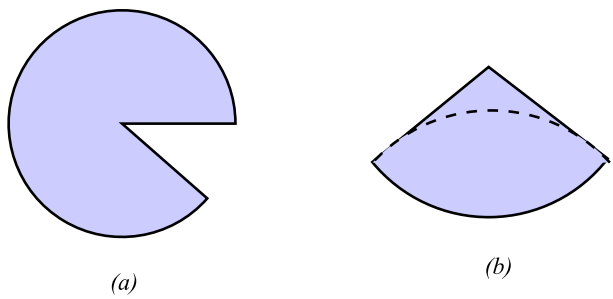}
   \end{center}
   \caption[C1]{In the figure, we see the region near the orbifold fixed point $\theta =\pi/2$. The deficit angle (a) is $\delta =2\pi (1-1/n)$.  After a rescaling, there is the conical singularity (b) at $\theta =\pi/2$.}
\label{fig:conical}
\end{figure}

Geometric entropy is an order parameter of the confinement/deconfinement phase transition for the  finite temperature Yang-Mills theory on the compactified space.
To describe the geometric entropy, we define the partition function on an orbifold $S^3/\mathbb{Z}_n$ to be $Z_{YM}(n)$: the ordinary partition function
 on $S^1\times S^3$ coincides with $Z_{YM}(1)$.
The metric of $S^3$ in the unit radius $R=1$ is given by
\ba &d\Omega_3^2=d\theta^2+\sin^2\theta d\phi^2+\cos^2\theta d\psi^2, \label{SPH21} \ea
where $0\le\theta \le\pi /2$ and $0\le\phi,\psi\le 2\pi$. We define the $\mathbb{Z}_n$ orbifold action as follows:
\ba \psi\sim \psi +\dfrac{2\pi}{n}.\ea
For $n\neq 1$, we have the conical singularities at $\theta =\pi/2$, with the deficit angle $\delta =2\pi (1-1/n)$ (see Figure. 1).  The submanifold defined by these singular points is $S^1$. 

Geometric entropy is described by
\ba S_G=-\dfrac{\partial}{\partial (1/n)}\log \left[\dfrac{Z_{YM}(n)}{Z_{YM}(1)^{\frac{1}{n}}}\right]_{n=1}. \ea
The geometric entropy is an analogue of the von-Neumann entropy as follows:
\ba  \dfrac{Z_{YM}(n)}{Z_{YM}(1)^{\frac{1}{n}}}=\mbox{Tr}\rho ^{\frac{1}{n}},\quad S_G=-\mbox{Tr}\rho\log\rho , \ea
where $\rho =e^{-2\pi H}$ is the density matrix when we regard the coordinate $\phi$ as the Euclidean time  and regard $H$ as its Hamiltonian. Note that the geometric entropy is not equal to von-Neumann entropy using the Boltzmann distribution in a canonical ensemble.

\section{Free energy of $d=4$ gauge theories on an orbifold $S^3/\mathbb{Z}_n$}
In this section, we compute the free energy of large $N$ $U(N)$ gauge theories with flavor on the orbifold $S^1\times S^3/\mathbb{Z}_n$ with the time direction compactified $(t\sim t+\beta)$ and with zero 't Hooft coupling limit $\lambda=0$. 
 We also calculate temperature dependence of the Polyakov loop vev in these theories. 
 
\subsection{Partition function}  
The matter contents of the $d=4$ $\mathcal{N}=2$ SYM are  $\mathcal{N}=2$ vector-multiplet and  $\mathcal{N}=2$ hyper-multiplet: the hyper-multiplet contains 4 real scalars and 2 Weyl fermions.
For free SYM theory on $S^3$, the matter field can be 
integrated out and the only remains are the temporal gauge field independent of positions. 
An exact expression for the partition function of the free $\mathcal{N}=2$ SYM on $S^1\times S^3$ is described by the following unitary matrix model \cite{schnitzer,Bas}:
\ba
&Z(v,f)= \nonumber \\
& \int [dU]\exp\Big[\sum ^{\infty}_{m=1}\dfrac{1}{m}\Bigl(
v(x^m)\mbox{Tr}U^m\mbox{Tr}U^{m \dagger} +\dfrac{1}{2}Nf(x^m)(\mbox{Tr}U^m+\mbox{Tr}U^{m\dagger})\Bigr)\Big], \label{P13} 
\ea
where $f=(N_f/N)f^{\prime}$ and $N_f$ is the number of the
fundamental-matters and $N_f/N=const$,
and the sum over $m$ denotes all the excitations of the single-particle
partition function of the oscillators for adjoint $v(x^m)$ and
fundamental representations $f'(x^m)$.

In the $\mathcal{N}=2$ SYM with flavor on the orbifold $S^3/\mathbb{Z}_n$, assuming $n$ is an odd integer,  single-particle partition function of the matters for adjoint and fundamental representation~\cite{taka,Hik} are given by
\ba 
&v(x^m)=v_B(x^m)+(-1)^{m+1}v_F(x^m), \label{SIN36}\\
&v_B(x)=2\,{\dfrac {  x ^{2} \left( 1+2\,
 x ^{n-1}- x  
  ^{n} \right) }{ \left( 1-x  \right) ^{2}
 \left( 1-x^{n} \right) }}+2\,{
\dfrac {x  \left( 1+ x  
 ^{n} \right) }{ \left( 1-x   \right) ^{2}
\left( 1-x^{n} \right) }}, \\
&v_F(x)=8\,{
\dfrac { x ^{n/2+1}}{ \left( 1-x
\right) ^{2} \left( 1- x  ^{n} \right) }}, 
\\&f^{\prime}(x^m)=f^{\prime}_B(x^m)+(-1)^{m+1}f^{\prime}_F(x^m), \label{SIN39} \\
& f_B^{\prime}(x)=4\,{\dfrac {x 
 \left( 1+x^{n} \right) }{ \left( 1
-x   \right) ^{2} \left( 1-x^{n} \right) }},\quad f_F^{\prime}(x)=8\,{\dfrac { x  ^{n/2+1}}{
 \left( 1-x  \right) ^{2} \left( 1-x^{n} \right) }}, \label{Sin11}
\ea
where $x=e^{-\beta}=e^{-1/T}$ at the temperature $T$ and 
the index $B,F$ means that we take the trace over the bosonic states and fermionic states, respectively. $T$ is defined in units of the inverse radius $1/R$ to be dimensionless.

We review these formulas on the orbifold 
$S^3/\mathbb{Z}_n$ following the appendix B of \cite{aharony}. Since the states of the field theory on $\mathbb{R}\times S^3$ are related with the local operators on $\mathbb{R}^4$ by a conformal transformation, the energy of the states is equal to the conformal dimension of the local operators. For this purpose, we embed the sphere $S^3$ in $\mathbb{C}^2$ with the coordinates $(z_1,z_2)$.
Then, we operate the following $\mathbb{Z}_n$ orbifold: 
\ba
z_1\sim e^{i\frac{2\pi}{n}}z_1.
\ea
For example, $\mathbb{Z}_n$ action operates on the scalar operator $\phi$ on $\mathbb{R}^4$ as follows:
\ba &\phi (z_1,\bar z_1,z_2,\bar z_2)\sim \phi (e^{i\frac{2\pi}{n}}z_1,e^{-i\frac{2\pi}{n}}\bar z_1,z_2,\bar z_2),\\
&\partial_1^2 \phi (z_1,\bar z_1,z_2,\bar z_2) \sim e^{i\frac{4\pi}{n}}\partial_1^2 \phi (e^{i\frac{2\pi}{n}}z_1,e^{-i\frac{2\pi}{n}}\bar z_1,z_2,\bar z_2),   \ea
where we defined the partial derivative $\partial_i=\partial /\partial z_i$ ($i=1,2$).

To obtain the single-particle partition function of the scalar fields, we sum the invariant operators under the $\mathbb{Z}_n$ action and the higher derivatives of the scalar operators with the traceless condition (the equation of motion). For $n=3$, we should count the following operators:
\ba
\phi ,\ \partial_2\phi ,\ \bar\partial_2 \phi ,\ \partial_2\partial_2 \phi, \ \partial_2\bar\partial_2 \phi ,\ \bar\partial_2\bar\partial_2 \phi ,\
\partial_1\partial_1\partial_1\phi,\ \bar\partial_1\bar\partial_1\bar\partial_1\phi ,...
\ea
Note that the local operator $...\partial_1\bar\partial _1\phi$ is 
 equivalent to $...\partial_2\bar\partial _2\phi$ because of the equation of motion.

The single-particle partition function is defined by
\ba
z(x)=\sum _{\text{local operators}}x^{\Delta}. \label{A14}
\ea
When we use the book keeping parameter $y$ and $y^{-1}$ for $\partial _1$ and $\bar\partial _1$, respectively,
 the partition function of a real boson \eqref{A14} becomes
\ba
&z(x)=\sum ^{\infty}_{k=1}kx^k+\sum^{\infty}_{k=2}x^k\sum ^{[(k-1)/n]}_{N=1}
\left(y^{nN}(k-nN)+y^{-nN}(k-nN)  \right)|_{y=1}, \no
&=\sum ^{\infty}_{k=1}kx^k(1+\sum^{\infty}_{l=1}\left(y^{nl}x^{nl}+y^{-nl}x^{nl}  \right))|_{y=1}, \no
&=\dfrac{x(1+x^n)}{(1-x)^2(1-x^n)}. 
\ea 

For the gauge fields $A_{\mu}$ on $\mathbb{R}^4$, the orbifold operates them as
\ba
&A_{z_1}(z_1,\bar{z}_1,z_2,\bar{z}_2)\sim e^{i\frac{2\pi}{n}}A_{z_1}(e^{i\frac{2\pi}{n}}z_1,e^{-i\frac{2\pi}{n}}\bar z_1,z_2,\bar z_2),  \\
&A_{z_2}(z_1,\bar{z}_1,z_2,\bar{z}_2)\sim A_{z_2}(e^{i\frac{2\pi}{n}}z_1,e^{-i\frac{2\pi}{n}}\bar z_1,z_2,\bar z_2).
\ea

The single-particle partition function for $A_{\mu}$ is obtained by summing
 the operators obeying the identification  $\partial_1\bar{\partial}_1A_{\mu}=-\partial _2\bar{\partial}_2A_{\mu}$ and satisfying
 the condition, 
 \ba
 &A^{\mu}=0,\ \partial_{(\mu_1}A_{\mu_2)}=0,\ \partial_{(\mu_1}\partial_{\mu_2}A_{\mu_3)}=0,....,
 \ea
 where () means symmetrization.
 These conditions are obtained by mapping the Gauss law constraint $A_{0}=0$ on $\mathbb{R}\times S^3$ to $x^{\mu}A_{\mu}=0$ on $\mathbb{R}^4$ and operating derivatives on $x^{\mu}A_{\mu}=0$.
 For $n=3$, we count the following invariant operators:
 \ba
& \partial_1A_{\bar{z}_1},\ \partial_2A_{\bar{z}_2},\ \partial_2\bar{\partial}_1A_{z_1},\ \bar{\partial}_2\bar{\partial}_1A_{z_1},\
\partial_2\bar{\partial}_2A_{z_2},\ \partial_2\bar{\partial}_2\bar{A}_{z_2}, 
 \ \partial_1^3A_{z_2},\ \bar{\partial}_1^3A_{z_2},...
 \ea
 The single-particle partition function becomes
\ba
&z_v(x)=2x\sum_{k=1}kx^k(1+2\sum_{l=1}x^{nl-1}) \nonumber \\
&=\dfrac{2x^2(1+2x^{n-1}-x^n)}{(1-x)^2(1-x^n)}.\label{VEC321}
\ea 

For Weyl fermions, the orbifold operates them including holonomy as 
\ba
\psi (z_1,\bar z_1,z_2,\bar z_2)\to e^{\frac{i\pi(n+1) }{n}\sigma_3}\psi (e^{i\frac{2\pi}{n}}z_1,e^{-i\frac{2\pi }{n}}\bar z_1,z_2,\bar z_2),
\ea
where Weyl fermions are of 2-component and operated by $\sigma_3$.
Note that the fermions are rotated by 1/2 of bosonic variables. Since every fermionic field is affected by the orbifold action, the invariant operators for $n=3$ are of the following form:
\ba
\partial_1\psi_1,\ \bar{\partial}_1\psi_2,\ \partial_2\partial_1\psi_1,\ \bar{\partial}_2\partial_1\psi_1,....
\ea
The single-particle partition function of a Weyl fermion becomes
\ba
&z_f=4x^{\frac{n}{2}}\sum_{k=1}kx^k(1+\sum_{l=1}x^{nl}) \nonumber \\
&=\dfrac{4x^{1+\frac{n}{2}}}{(1-x)^2(1-x^n)}, \label{PAR224}
\ea
where 4 in the numerator of \eqref{PAR224} comes from 
 contributions of complex fermions $\psi_1$ and $\psi_2$.

\subsection{Free energy}

We derive the free energy of $\mathcal{N}=2$ SYM by using the approximation $v(x^m)=f(x^m)=0$ $(m>1)$. Namely, the first winding state in the time direction is only excited. This approximation is valid for not sufficiently high temperature region.
By introducing the Lagrange multiplier $\lambda,\bar\lambda$~\cite{hliu,wadia}, \eqref{P13} is rewritten as
\ba
&Z(v,f)=\dfrac{N^2}{8\pi v}\int [dU]d\lambda \bar d\lambda
\exp \left[-\dfrac{N^2}{4v}(\lambda-f)(\bar\lambda-f)+\dfrac{N}{2}
(\lambda\mbox{Tr}U+\bar\lambda \mbox{Tr}U^{\dagger})  \right] \\
&=\dfrac{N^2}{4\pi v}\int^{\infty}_0gdg \int ^{\pi}_{-\pi}d\theta
\exp \left[-\dfrac{N^2}{4v}(g^2-2gf\cos\theta +f^2 )\right]\cdot \\
&\cdot\int [dU]\exp\left(\dfrac{Ng}{2}(\mbox{Tr}U+\mbox{Tr}U^{\dagger})\right) \\
&=\dfrac{N^2}{2v}\int^{\infty}_0gdge^{-N^2\beta F(v,f,g)},
\ea
where
\ba N^2\beta F(v,f,g)=-\log I_0\left(\dfrac{N^2gf}{2v}\right)+\dfrac{N^2}{4v}(f^2+g^2)-N^2K(g) \label{F210}  \ea
and we defined 
\ba e^{N^2K(g)}=\int [dU]\exp{\dfrac{1}{2}Ng\left(\mbox{Tr}U+\mbox{Tr}U^{\dagger}\right)}.  \ea
The asymptotic expansion of $K(g)$ in the large $N$ limit is computed in 
\cite{hliu} as follows:
\ba K(g)=\begin{cases} \dfrac{g^2}{4}+O(1/N^3) & (g<1) \\
g-\dfrac{1}{2}\log g-\dfrac{3}{4}+O(1/N^2) & (g>1) \end{cases}  \ea
We also expand the first term in \eqref{F210} in powers of $N^2$ as follows:
\ba
\log I_0\left(\dfrac{N^2gf}{2v} \right) =\dfrac{N^2gf}{2v}
-\dfrac{1}{2}\log \left( \dfrac{\pi N^2gf}{v}\right)-\dfrac{v}{4N^2gf}
+O(N^{-4})
\ea

\begin{figure}[htbp]
   \begin{center}
     \includegraphics[height=6cm]{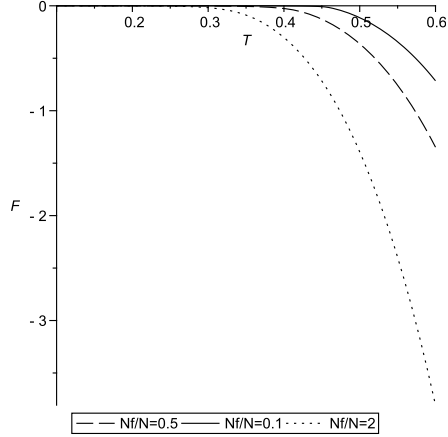}
   \end{center}
   \caption[aa]{Temperature dependence of the free energy for the $\mathcal{N}=2$ SYM with flavor is plotted.}
\label{Free}
\end{figure}
We analyze the phase structure by using the saddle point approximation
and by looking at the leading order terms in \eqref{F210} in powers of $N^2$.

When $v<1$ and $f <f_0=1-v$, $F(v,f,g)$ has a minimum at
\ba
g_0=\dfrac{f}{f_0}<1,\quad \beta F(v,f,g_0)=-\dfrac{f^2}{4(1-v)}. \label{F218}
\ea
This behavior of free energy implies that the fermions are not the fundamental degrees of freedom but the mesons are
fundamental degree of freedom $N^2F\propto N_f^2$ at the low temperature.

When $v< 1$ and $f>f_0$ or $v>1$, $F(v,f,g)$ has a minimum at 
\ba
g_0=v +\dfrac{f}{2}+\sqrt { \left( {v}+\dfrac{f}{2} \right) ^{2}-{v}},\quad
 \beta F(v,f,g_0)=-\dfrac{g_0}{2}-{\frac {f{g_0}}{4v}}+\dfrac{1}{2}+\dfrac{1}{2}\log g_0  +{\frac { f^{2}}{4{v}}}. \label{F219}
\ea
We plot $T$-dependence of the free energy for 
various $N_f$ in Fig.~\ref{Free}.

Note that the critical line of the above phases is described by the following formula:
\ba
v(x)+f(x)=1,\quad g_0=1. \label{C220}
\ea

As was shown in the literature \cite{hliu,schnitzer}, there is the Gross-Witten type third order phase transition at the temperature $T=T_c$ determined by the formula \eqref{C220}.

\subsection{Polyakov loop vev}
\begin{figure}[htbp]
   \begin{center}
     \includegraphics[height=6cm]{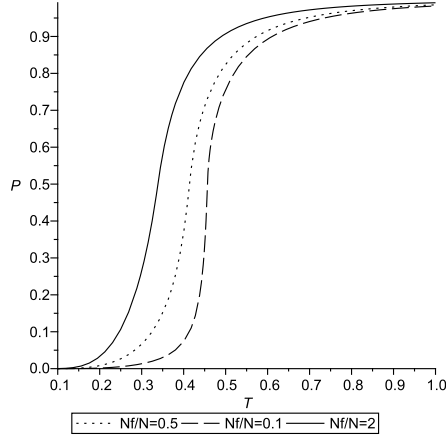}
   \end{center}
   \caption[aa]{Temperature dependence of Polyakov loop vev is plotted. $T$ is in units of $1/R$. The critical temperature $T_c$ at which $P=1/2$ is equal to $0.46,0.41,0.34$ for $N_f/N=0.1,0.5,2$, respectively. }
\label{Pol}
\end{figure}
It is interesting to obtain the Polyakov loop expectation value.
According to~\cite{gross}, the density of eigenvalues becomes in our notation,
\ba
\rho =\begin{cases}
&\dfrac{g}{\pi}\cos\dfrac{\alpha}{2}\Big(\dfrac{1}{g}-\sin^2\dfrac{\alpha}{2}\Big)^{\frac{1}{2}} \quad g\ge 1, \quad \sin^2\frac{\alpha}{2}\le \frac{1}{g}, \\
&\dfrac{1}{2\pi}(1+g\cos\alpha) \quad g\le 1, \quad |\alpha|\le \pi.  \\
\end{cases} \label{Rho336}
\ea
Polyakov loop vev becomes the first moment of $\rho$ as follows: 
\ba
P=\rho_1=\int ^{\pi}_{-\pi}\rho\cos (\alpha)d\alpha.
\ea
And it can be obtained from \eqref{Rho336} as
\ba
\rho_1 =\begin{cases}
&1-\frac{1}{2(v +f/2+\sqrt { \left( {v}+\frac{f}{2} \right) ^{2}-{v}})} \quad g\ge 1, \\
&\dfrac{f}{2(1-v)} \quad g\le 1. \\
\end{cases}
\ea
We plotted temperature dependence of Polyakov loop vev for $N_f/N=2,0.5,0.1$ in Fig.~\ref{Pol}. It is interesting to compare Fig. \ref{Pol} with the Figure in~\cite{And}. Our results of Polyakov loop vev have a similar behavior with the lattice results and the soft-wall AdS/QCD results written in~\cite{And}.

\section{Geometric entropy of $\mathcal{N}=2$ SYM}
\begin{figure}[htbp]
   \begin{center}
     \includegraphics[height=6cm]{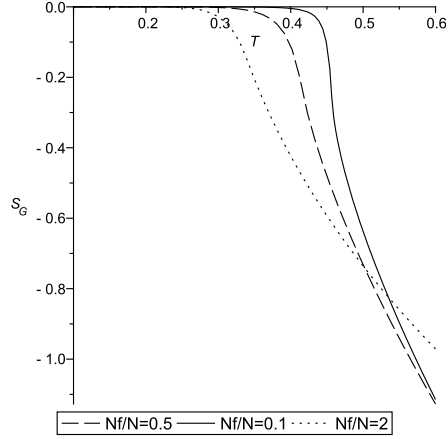}
   \end{center}
   \caption[ Geometric entropy]{Temperature dependence of geometric entropy is plotted. $T$ is in units of $1/R$.}
\label{fig:geo}
\end{figure}
\begin{figure}[htbp]
   \begin{center}
     \includegraphics[height=6cm]{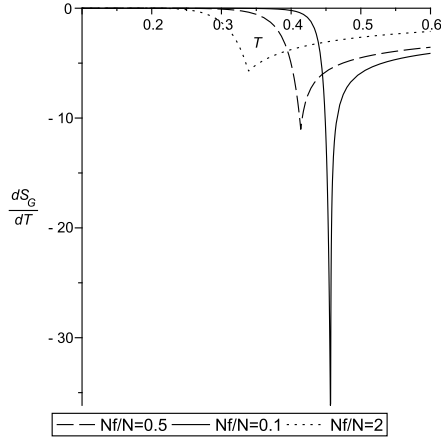}
   \end{center}
   \caption[1-derivative of geometric entropy]{Temperature dependence of $dS^{G}/dT$ is plotted. $T$ is in units of $1/R$. The phase transition happens at $T_c=0.46$, $0.41$, $0.34$, for $N_f/N=0.1$, $0.5$, $2$, respectively.}
\label{fig:geo2}
\end{figure}
By using the free energy \eqref{F218} and \eqref{F219}, we can describe geometric entropy of $\mathcal{N}=2$ SYM as follows:
\ba
&S_G=-\dfrac{\partial}{\partial (1/n)}\left( \log Z(n)-\dfrac{1}{n}\log Z(1)\right)\Biggr|_{n=1} \no
&=-\dfrac{\partial}{\partial n}\left(( \beta F)(n)-\dfrac{1}{n}(\beta F(1))\right). \label{G221}
\ea
where $Z(n)\equiv Z(v,f)$ and $F(n)\equiv F(v,f,g_0)$.
Computing \eqref{G221}, we plot $S_G$ and $dS_G/dT$ near the critical temperature in Figure.~\ref{fig:geo} and Figure.~\ref{fig:geo2}, respectively. We find that since geometric entropy 
is a kind of von-Neumann entropy,
 geometric entropy of $\mathcal{N}=2$ SYM can capture the third order phase transition of the Gross-Witten matrix model and the critical temperature $T_c$ decreases if $N_f/N$ increases. $T_c$ is consistent with the value obtained in \eqref{C220}. 
 
It will be interesting to consider the case at non-zero 't Hooft coupling. According to~\cite{Bas}, the order of the phase transition will become the first order at non-zero 't Hooft couping. So, it is expected that in this case, the geometric entropy can capture the first order confinement/deconfinement transition. 
 \subsection{Analysis for the high temperature limit}
In the high temperature limit $x\to 1$ $(\beta \to 0)$, the single-particle partition functions on $S^3/\mathbb{Z}_n$~\eqref{VEC321} and~\eqref{PAR224}  are given by

\ba
z_{c}(x^m)=\dfrac{4}{m^3n\beta^3}+\dfrac{n^2-1}{3mn\beta}+O(\beta), \nonumber \\
z_v(x^m)=\dfrac{4}{m^3n\beta^3}+\dfrac{n^2-6n-1}{3mn\beta}+O(1), \nonumber \\
z_f(x^m)=\dfrac{4}{m^3n\beta^3}-\dfrac{2+n^2}{6mn\beta}+O(\beta). \label{SIN03}
\ea
where $z_c(x)$ is the single-particle partition function of a complex boson.
Using \eqref{SIN03}, the free energy and the geometric entropy of the $\mathcal{N}=2$ SYM are obtained as follows (see \cite{aharony}):
\ba
&F=-\dfrac{1}{12}\pi^2 N^2T^4\left(1+\dfrac{N_f}{N}\right) V_{S^3}, \\
&S^{}_G=\dfrac{\pi^2 N^2}{6\beta}\left( \dfrac{N_f}{N}-1 \right), \label{Geo223}
\ea
where $V_{S^3}=2\pi^2$.
The above result implies that the geometric entropy becomes zero similar to the case of $\mathcal{N}=4$ SYM if $N=N_f$ and the geometric entropy changes its sign depending on $N$ and $N_f$.\footnote{The coefficient in~\eqref{Geo223} is similar to the $\beta$ function of $\mathcal{N}=2$ SYM theory. However, in the QCD case with no supersymmetry, the geometric entropy has minus value for any $N_f$ since the only positive contribution to the geometric entropy is obtained from that of scalars and periodic fermions (see~\cite{taka}).}  

According to~\cite{taka},  $S_G$ should be subtracted by the geometric entropy $S^P_G$, where $S^P_G$ is geometric entropy with the fermions obeying the periodic boundary condition in the time direction: $\mbox{tr}(-1)^Fe^{-\beta H}$ instead of $\mbox{tr}e^{-\beta H}$.  The partition function with periodic fermions is easily obtained by removing $(-1)^{n}$ in \eqref{SIN36} and \eqref{SIN39}. Thus, $S^P_G$ and $\Delta S_G= S_G-S^P_G$ is obtained as follows:
\ba
&S^P_G=\dfrac{\pi^2 NN_f}{3\beta}, \\
&\Delta S_G=-\dfrac{\pi^2 N^2}{6\beta}\left( \dfrac{N_f}{N}+1 \right). \label{DS225}
\ea 
The result \eqref{DS225} is plausible since the coefficient of $\Delta S_G$ is proportional to that of the entropy $-dF/dT$.
 
We want to comment on the relation to the gauge/gravity correspondence for $N_f/N\ll 1$. $\Delta S_G$ seems to be comparable with the dual gravity result in the high temperature limit.  
However, the analysis in the dual gravity side seems to be difficult since in the probe D7-brane analysis~\cite{Karc,Mate}, geometric entropy has no flavor contributions. It implies that higher order correction on the D7-brane worldvolume and the background charges of the D7-brane should be included.
 
\section{Introducing chemical potential}

In this section, we analyze the phase structure of $d=4$ $\mathcal{N}=2$ SYM with flavor at finite chemical potential. The similar analysis in $d=4$ $\mathcal{N}=4$ SYM
 was done in~\cite{Bas2,Yam,Mur}. 
 We start with reviewing global symmetry of $d=4$ $\mathcal{N}=2$ SYM with flavor. The form of the Lagrangian for $N_f$ hypermultiplets $(Q_a(q,\psi_q),\tilde{Q}_a(\tilde{q},\tilde{\psi}_q))$ (labeled by an index $a$), interacting with a $\mathcal{N}=2$ vector multiplet is given by
\ba
\mathcal{L}=\int (Q_a^{\dagger}e^{-2V}Q_a+\tilde{Q}_ae^{2V}\tilde{Q}^{\dagger}_a)+\int d^2\theta (\tilde{Q}_a\Phi Q_a)+h.c.\label{Lag541}
\ea
where $V$ is the $\mathcal{N}=1$ superfield and $\Phi$ is chiral superfield in the adjoint representation. In $\mathcal{N}=2$ SYM theory, there are $U(1)_J$ and $U(1)_R$ subgroups\footnote{Remind that $U(1)_R$ operates on $\mathcal{N}=2$ fields as a chiral symmetry that is quantum-mechanically broken to the discrete $Z_{4N-2N_f}$ due to anomalies except the case $N_f=2N$. }
 in $SU(2)_R$ $R$-symmetry~\cite{Alv}. In this paper, we don't consider chemical potential conjugate to $U(1)_R$ since $\mathcal{N}=2$ SYM with flavor on $S^1\times S^3$ will be not free from the anomaly for large $N_f$.
The transformations of $\mathcal{N}=1$ superfields under $U(1)_J$ are given by
\ba
U(1)_J:
&\Phi\to \Phi(e^{-i\alpha}\theta),\quad V\to V(e^{-i\alpha}\theta), \\
&Q\to e^{i\alpha}Q(e^{-i\alpha}\theta),\quad \tilde{Q}\to e^{i\alpha}\tilde{Q}(e^{-i\alpha}\theta),
\ea
And the transformations of all component fields under $U(1)_J$ are given by
\ba
&(\lambda ,q)\to e^{i\alpha}(\lambda,q), \nonumber \\
&(\psi,\tilde{q}^{\dagger}) \to e^{-i\alpha}(\psi,\tilde{q}^{\dagger}), 
\ea
where $(\psi,\lambda)$ are fermions in the $\mathcal{N}=2$ vector multiplet. 

In addition, there is an $U(1)_F$ subgroup in $U(N_f)$ flavor symmetry as follows:
\ba
(Q_a,\tilde{Q}_a^{\dagger})\to e^{i\alpha}(Q_a,\tilde{Q}_a^{\dagger}).
\ea
$U(1)_F$ symmetry is baryonic symmetry of $\mathcal{N}=2$ SYM with flavor.

In the next section and the appendix A, we introduce chemical potential conjugate to these $U(1)$ symmetry and analyze the phase structure of $\mathcal{N}=2$ SYM at finite density.

\subsection{$U(1)_J$ case}
\begin{figure}[!htb]
   \begin{center}
     \includegraphics[height=6cm]{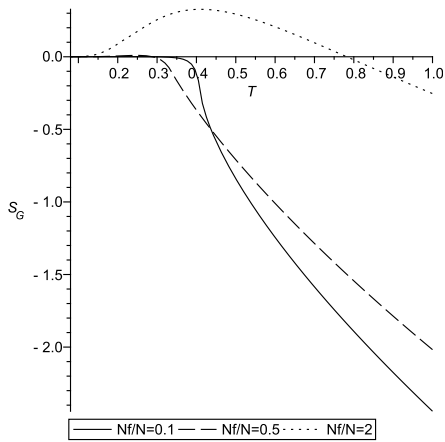}
   \end{center}
   \caption[]{Temperature dependence of geometric entropy is plotted at finite chemical potential $\mu=1/2$. In the figure, $T$ and $\mu$ are measured in units of $1/R$. Our plot shows that $S_G$ at low temperature and for large $N_f$ takes the positive value.}
\label{SGJ}
\end{figure}
\begin{figure}[!htb]
   \begin{center}
     \includegraphics[height=6cm]{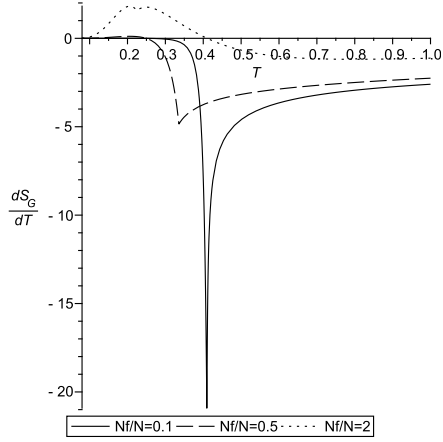}
   \end{center}
   \caption[]{Temperature dependence of $dS_G/dT$ is plotted at finite chemical potential $\mu=1/2$. $T$ and $\mu$ are in units of $1/R$. The phase transition happens at $T_c=0.41$, $0.34$, and $0.22$ for $N_f/N=0.1$, $0.5$, and $2$, respectively.}
\label{dSdTJ}
\end{figure}

\begin{figure}[!htb]
   \begin{center}
     \includegraphics[height=6cm]{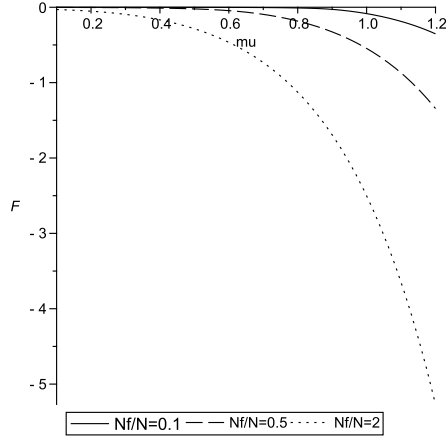}
   \end{center}
   \caption[]{$\mu$ dependence of $F$ at $T=0.3$ is plotted. $\mu$ is measured in units of $1/R$.}
   \label{FmJ}
   \end{figure}
\begin{figure}[!htb]
   \begin{center}
     \includegraphics[height=6cm]{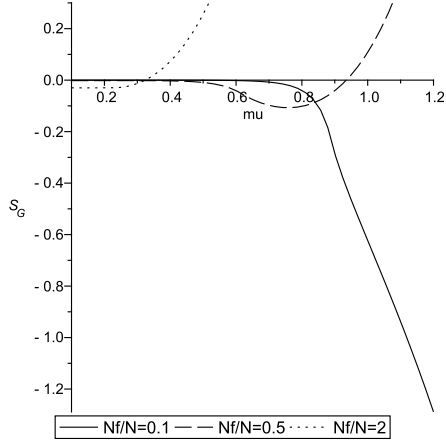}
   \end{center}
   \caption[ ]{$\mu$ dependence of $S_G$ at $T=0.3$ is plotted. $\mu$ is in units of $1/R$. Our plot shows that $S_G$ at low temperature and for large $N_f$ takes the positive value as $\mu$ becomes large.}
\label{GmuJ}
\end{figure}
 \begin{figure}[!htb]
   \begin{center}
     \includegraphics[height=6cm]{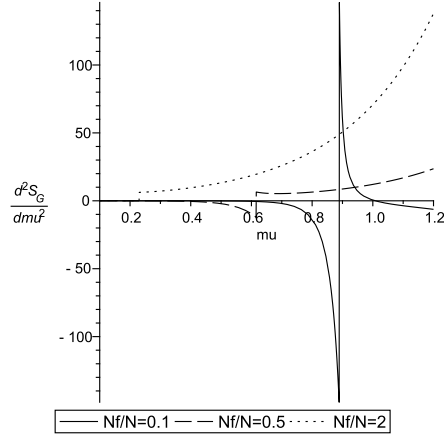}
   \end{center}
   \caption[]{$\mu$ dependence of $d^2S_G/d\mu^2$ at $T=0.3$ is plotted. $\mu$ is in units of $1/R$. The phase transition happens at the critical values of chemical potential $\mu_c=0.89$, $0.62$, and $0.23$ for $N_f/N=0.1$, $0.5$, and $2$, respectively.}
\label{ddGmuJ}
\end{figure}

    \begin{figure}[!htb]
   \begin{center}
     \includegraphics[height=6cm]{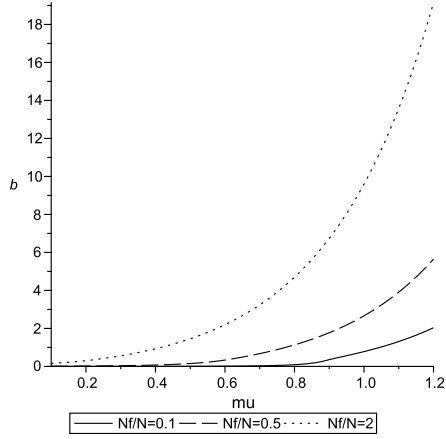}
   \end{center}
   \caption[]{$\mu$ dependence of the number density $b$ is plotted at finite density and $T=0.3$. $\mu$ is in units of $1/R$. $b$ increases monotonously as $\mu$ increases as expected.}
\label{bmuJ}
\end{figure}
 \begin{figure}[!htb]
   \begin{center}
     \includegraphics[height=6cm]{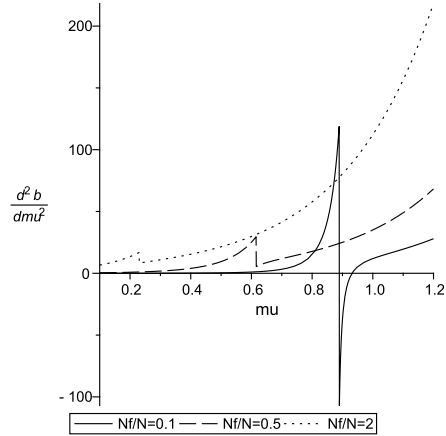}
   \end{center}
   \caption[]{$\mu$ dependence of $d^2b/d\mu^2$ is plotted at finite density and $T=0.3$. $\mu$ is in units of $1/R$. }
\label{ddbmuJ}
\end{figure}
For $U(1)_J$ case, we make the approximation of ignoring the higher order of winding states $m>1$. This approximation is valid for not high temperature and not at large density\footnote{Note that there are also stability bounds $E\ge \mu Q$, where $E$ is the energy of particle states and $Q$ is an $U(1)$ charge of these states: in \eqref{BRR1}, the bound is $\mu <1$.}. Then, the phase structure of the system at finite chemical potential can be analyzed by evaluating free energy, geometric entropy, and number density. 

For single-particle partition functions including chemical potential $\mu$ conjugate to the $U(1)_J$ charge, we should replace \eqref{SIN36}-\eqref{Sin11} by
 \ba
 &v(x^m)=v_B(x^m)+(-1)^{m+1}v_F(x^m), \\
&v_B(x)=2\,{\dfrac {  x ^{2} \left( 1+2\,
 x ^{n-1}- x^{n} \right) }{ \left( 1-x  \right) ^{2}
 \left( 1-x^{n} \right) }}+2\,{
\dfrac {x  \left( 1+ x  
 ^{n} \right) }{ \left( 1-x   \right) ^{2}
\left( 1-x^{n} \right) }}, \ea \ba &v_F(x)=4\,{
\dfrac { x ^{n/2+1+\mu}}{ \left( 1-x
\right) ^{2} \left( 1- x  ^{n} \right) }}+4\,{
\dfrac { x ^{n/2+1-\mu}}{ \left( 1-x
\right) ^{2} \left( 1- x  ^{n} \right) }}, \\
&f^{\prime}(x^m)=f^{\prime}_B(x^m)+(-1)^{m+1}f^{\prime}_F(x^m), \\
& f_B^{\prime}(x)=4\,{\dfrac {x^{1-\mu} 
 \left( 1+x^{n} \right) }{ \left( 1
-x   \right) ^{2} \left( 1-x^{n} \right) }},\quad f_F^{\prime}(x)=8\,{\dfrac { x  ^{n/2+1}}{
 \left( 1-x  \right) ^{2} \left( 1-x^{n} \right) }}. \label{BRR1}
\ea
Here, we introduce the number density $b$ depending on $\mu$ as an order parameter of the third order phase transition. We define $b$ as
\ba
b=-\dfrac{dF}{d\mu}.
\ea

The free energy can be evaluated by calculating \eqref{F218} and \eqref{F219} in the presence of chemical potential and has behavior similar to $\mu=0$ for $\mu\neq 0$.  We plotted temperature dependence of $S_G$ at $\mu=1/2$ and $dS_G/dT$ at $\mu=1/2$ in Fig.~\ref{SGJ} and Fig.~\ref{dSdTJ}, respectively. We observe the third order phase transition in the presence of finite chemical potential $\mu=1/2$. 

The geometric entropy for the large flavor $N_f/N\ge 1$ has interesting behavior at low temperature. For the large flavor $N_f/N=2$, the geometric entropy takes the larger positive value at low temperature than $N_f/N=0.1$, $0.5$ cases. The similar behavior is also observed evaluating $S_G$ and $dS_G/d\mu$ as the functions of $\mu$ in Fig.~\ref{GmuJ} and Fig.~\ref{ddGmuJ}. 
These behavior of the geometric entropy at the low temperature seems to come from the relation to the scale anomaly that is proportional to the central charges. Since the definition of the geometric entropy includes changes of the area Area$(S^3/\mathbb{Z}_n)$ by varying the orbifold action, the geometric entropy at the low temperature will be related with a scale anomaly of $\mathcal{N}=2$ SYM theory on $\mathbb{R}\times S^3$.\footnote{The period of $S^1$ is large at low temperature: $S^1\simeq  \mathbb{R}$. }

Moreover, $\mu$ dependence of $b$ and $d^2b/d\mu^2$ at $T=0.3$ is plotted in Fig.~\ref{bmuJ} and Fig.~\ref{ddbmuJ}, respectively. Our results show that both $b$ and $S_G$ are order parameters of the third order phase transition.

\section{Discussion}
In this paper, we investigated the phase structure of $d=4$  $\mathcal{N}=2$ large $N$ SYM theory with flavor at finite temperature by using geometric entropy as an order parameter. Analyzing geometric entropy, namely von-Neumann entropy associated with the space translation, we observed the third order phase transition as the confinement/deconfinement transition instead of the first order phase transition. We also enlarged our analysis to the finite density system and observed the third order phase transition in respect to $T$ or $\mu$. The geometric entropy at finite chemical potential for large $N_f$ has the interesting behavior at low temperature implying that the scale anomaly of $\mathcal{N}=2$ SYM on $\mathbb{R}\times S^3$ may contribute. 
    
In the appendix A, we also analyzed the system in the presence of the baryon chemical potential conjugate to $U(1)_F$ in the low temperature limit since the similar analysis at finite temperature is more complicated. We found the confinement phase similar to the quark matter phase in the large $N$ QCD at the finite chemical potential $\mu>1$. For future works, we should include the interaction terms of bosons to make the baryon number density with positive chemical potential $\mu>0$ well defined.     
    
It will be interesting to include mass terms for the flavor field as had been done in~\cite{Han} for large $N$ and $N=3$ QCD theory since the $(T,\mu)$ phase space of two theories has different behavior. Moreover, it is important to compute the chiral condensate of $\mathcal{N}=2$ SYM theory with massive flavor fields and to argue the chiral symmetry breaking of this theory.  
It will also be interesting to enlarge our analysis for different orbifold systems~\cite{Hik} since geometric entropy depends on the orbifold singularity. 

\vskip2mm
\noindent {\bf Acknowledgments}: 
M. F. would like to thank theoretical particle physics members in Kyoto U., K. Fukushima, Y. Hidaka, T. Ishii, T. Nishioka, and T. Takayanagi for the discussions and helpful comments. H. O. would like to thank Joyce C. Myers for fruitful discussions and comments. M. F. would like to thank T. Tai for careful reading of this manuscript. M. F. is supported in part by the Japan Society for the Promotion of Science. H. O. is supported in part by the Grant-in-Aid for Scientific Research No. 21-897 from the Ministry of Education, Culture, Sports, Science and Technology of Japan.

\vskip2mm

\appendix

\section{$U(1)_F$ case and low temperature limit}
$U(1)_F$ symmetry is similar to the baryonic symmetry in QCD theories. For the $U(1)_F$ case, it seems to be interesting to investigate finite density and the low temperature phase of $SU(N)$ gauge theories such as quark matter phase in the large $N$ QCD~\cite{McL}, while the analysis at finite temperature by using geometric entropy is complicated. In this subsection,  the $(T,\mu)$ phase space near $\mu = 1$ and at low temperature is investigated by using saddle point approximation.
Introducing chemical potential conjugate to $U(1)_F$, the chiral field $Q_a$ in the $\mathcal{N}=2$ hypermultiplet survives at low temperature ($R/\beta \ll 1$)~\cite{Han}. In addition, fermionic modes and higher order of scalar modes can be ignored in the small sphere limit. In the low temperature limit, the partition function becomes
\ba
&Z=\int \prod_i[d\theta_i]\exp(-S(\theta_i)), \\ 
&S(\theta_i)=-\dfrac{1}{2}\sum_{ij}\log \sin^2\Big(\dfrac{\theta_i-\theta_j}{2}\Big)+N\sum_iV(\theta_i), \\
&V(\theta)=
\sigma_1 \log(1-\xi e^{i\theta})+i\mathcal{N}\theta, \label{VTH557}
\ea
where $\xi=e^{\beta(\mu-1)}$ and we introduced the Lagrange multiplier $\mathcal{N}$ enforcing the $ \sum_i\theta_i=0$ constraint and focused on the lowest energy level of the bosonic states, $\sigma_1=2N_f/N$.\footnote{The generalization to the multi-level model is straightforward. However, it will be more complicated to find the boundary in the $(\mu,T)$ plane.}

Introducing the parameter $s$ satisfying $-\pi<s<\pi$ and $s\in \mathbb{R}$ and defining $z_i=\exp(i\theta_i)$,
the eigenvalue distribution in the large $N$ limit is defined by
\ba
i\dfrac{ds}{dz}=\rho(z).
\ea 
 The saddle point equation thus becomes
\ba
zV'(z)=P\oint \dfrac{dz'}{2\pi i}\rho(z')\dfrac{z+z'}{z-z'},\quad zV'(z)=\mathcal{N}-\dfrac{\sigma_1\xi z}{1-\xi z}. \label{SAD559}
\ea 
$\rho(z)$ has poles at $z=0$ and $1/\xi$. There are two cases where the pole at $1/\xi$ is outside the contour of integral or not. Note that these two cases are related with each other by the transformation $\xi\to 1/\xi$. So, it is enough to consider the small $\xi$ case where $1/\xi$ is outside the contour.
In the small $\xi$ limit, the pole at $1/\xi$ is outside the contour of integral. Then, the following $\rho(z)$ satisfies \eqref{SAD559}: 
\ba
\rho(z)=\dfrac{1}{z}+\dfrac{\sigma_1 \xi}{1-\xi z}.
 \ea
Polyakov loop vev is given by
\ba
P_1=\oint \dfrac{dz}{2\pi i}\rho(z)z=0.
\ea 
Vanishing Polyakov vev implies that the confinement happens.
The phase transitions to the deconfinement phase happen at the place where $\rho(z)$ vanishes precisely at the point $z(\pm \pi)$.
The line of the phase transitions in the $(\mu,T)$ plane is given by
\ba
\mu=1+T((\sigma_1-1)\log(-1+\sigma_1)-\sigma_1\log\sigma_1). \label{MUT562}
\ea
To make \eqref{MUT562} well defined, the condition $N_f/N\ge 1/2$ should be satisfied. And the confinement region at finite chemical potential exists if $N_f/N>1/2$ implying that the confinement phase of $\mathcal{N}=2$ SQCD is derived  at least by evaluating the fermionic states. The next confinement region is separated by the line 
\ba
\mu=1-T((\sigma_1-1)\log(-1+\sigma_1)-\sigma_1\log\sigma_1). \label{MUT563}
\ea
The above line is obtained by replacing $\xi$ with $1/\xi$ in \eqref{VTH557}. Note that the line~\eqref{MUT562} and~\eqref{MUT563} reach $\mu=1$ as $N_f/N\to 1/2$. And the deconfinement phase vanishes at the low temperature.  
This behavior is similar to that of the large $N$ QCD in a small $S^3$ analyzed in~\cite{Han}. For these cases, it can be shown that in the small $S^3$ limit and as $N_f/N\to 0$, the corresponding matrix model can realize the phase structure of the large $N$ QCD in the low temperature limit!\footnote{According to~\cite{Hid}, furthermore, the baryon degeneracy at large $N$ has the function similar to the coefficient of $T$ in \eqref{MUT562}.} This phase structure contains the quark matter phase where the pressure behaves as $O(N)$ and the Polyakov loop vev vanishes.


\vskip2mm








\begin{thebibliography}{99}
\bibitem{taka}
  M.~Fujita, T.~Nishioka and T.~Takayanagi,
  ``Geometric Entropy and Hagedorn/Deconfinement Transition,''
  JHEP {\bf 0809}, 016 (2008)
  [arXiv:0806.3118 [hep-th]].
\bibitem{HLW}
  C.~Holzhey, F.~Larsen and F.~Wilczek,
  ``Geometric and renormalized entropy in conformal field theory,''
  Nucl.\ Phys.\  B {\bf 424}, 443 (1994)
  [arXiv:hep-th/9403108].

\bibitem{CC}
  P.~Calabrese and J.~L.~Cardy,
  ``Entanglement entropy and quantum field theory,''
  J.\ Stat.\ Mech.\  {\bf 0406}, P002 (2004)
  [arXiv:hep-th/0405152].


\bibitem{Area}
  L.~Bombelli, R.~K.~Koul, J.~H.~Lee and R.~D.~Sorkin,
  ``A Quantum Source of Entropy for Black Holes,''
  Phys.\ Rev.\  D {\bf 34}, 373 (1986);
 M.~Srednicki,
  ``Entropy and area,''
  Phys.\ Rev.\ Lett.\  {\bf 71}, 666 (1993)
  [arXiv:hep-th/9303048].

\bibitem{ma}
  J.~M.~Maldacena,
  { The large N limit of superconformal field theories and supergravity,}
  Adv.\ Theor.\ Math.\ Phys.\  {\bf 2}, 231 (1998)
  [Int.\ J.\ Theor.\ Phys.\  {\bf 38}, 1113 (1999)]
  [arXiv:hep-th/9711200].

\bibitem{gkp}
  S.~S.~Gubser, I.~R.~Klebanov and A.~M.~Polyakov,
  { Gauge theory correlators from non-critical string theory,}
  Phys.\ Lett.\  B {\bf 428}, 105 (1998)
  [arXiv:hep-th/9802109].

\bibitem{wi}
  E.~Witten,
  { Anti-de Sitter space and holography,}
  Adv.\ Theor.\ Math.\ Phys.\  {\bf 2}, 253 (1998)
  [arXiv:hep-th/9802150].
  

\bibitem{sundberg}
  B.~Sundborg,
  { The Hagedorn transition, deconfinement and N = 4 SYM theory,}
  Nucl.\ Phys.\  B {\bf 573}, 349 (2000)
  [arXiv:hep-th/9908001].


\bibitem{aharony}
  O.~Aharony, J.~Marsano, S.~Minwalla, K.~Papadodimas and M.~Van Raamsdonk,
  { The Hagedorn / deconfinement phase transition in weakly coupled large N
  gauge theories,}
  Adv.\ Theor.\ Math.\ Phys.\  {\bf 8}, 603 (2004)
  [arXiv:hep-th/0310285].
\bibitem{Aha2}
  O.~Aharony, J.~Marsano, S.~Minwalla, K.~Papadodimas and M.~Van Raamsdonk,
  ``A first order deconfinement transition in large N Yang-Mills theory on a
  small 3-sphere,''
  Phys.\ Rev.\  D {\bf 71}, 125018 (2005)
  [arXiv:hep-th/0502149].

\bibitem{Kar2}
  A.~Karch, E.~Katz, D.~T.~Son and M.~A.~Stephanov,
  ``Linear Confinement and AdS/QCD,''
  Phys.\ Rev.\  D {\bf 74}, 015005 (2006)
  [arXiv:hep-ph/0602229].
\bibitem{And}
  O.~Andreev,
  ``Renormalized Polyakov Loop in the Deconfined Phase of SU(N) Gauge Theory
  and Gauge/String Duality,''
  Phys.\ Rev.\ Lett.\  {\bf 102}, 212001 (2009)
  [arXiv:0903.4375 [hep-ph]].

\bibitem{Fujj2}
  M.~Fujita, K.~Fukushima, T.~Misumi and M.~Murata,
  ``Finite-temperature spectral function of the vector mesons in an AdS/QCD
  model,''
  Phys.\ Rev.\  D {\bf 80}, 035001 (2009)
  [arXiv:0903.2316 [hep-ph]].


\bibitem{Fujjj2}
  M.~Fujita, T.~Kikuchi, K.~Fukushima, T.~Misumi and M.~Murata,
  ``Melting Spectral Functions of the Scalar and Vector Mesons in a Holographic
  QCD Model,''
  Phys.\ Rev.\  D {\bf 81}, 065024 (2010)
  [arXiv:0911.2298 [hep-ph]].





\bibitem{RT}
 S.~Ryu and T.~Takayanagi,
  ``Holographic derivation of entanglement entropy from AdS/CFT,''
  Phys.\ Rev.\ Lett.\  {\bf 96} (2006) 181602
  [arXiv:hep-th/0603001];
``Aspects of holographic entanglement entropy,''
  JHEP {\bf 0608} (2006) 045
  [arXiv:hep-th/0605073];

\bibitem{Nishioka:2006gr}
  T.~Nishioka and T.~Takayanagi,
  ``AdS bubbles, entropy and closed string tachyons,''
  JHEP {\bf 0701}, 090 (2007)
  [arXiv:hep-th/0611035].

\bibitem{Nis}
  T.~Nishioka, S.~Ryu and T.~Takayanagi,
  ``Holographic Entanglement Entropy: An Overview,''
  J.\ Phys.\ A  {\bf 42}, 504008 (2009)
  [arXiv:0905.0932 [hep-th]].


\bibitem{hliu}
  H.~Liu,
  ``Fine structure of Hagedorn transitions,''
  arXiv:hep-th/0408001.
    \bibitem{gross}
  D.~J.~Gross and E.~Witten,
  { Possible Third Order Phase Transition In The Large N Lattice Gauge
  Theory,}
  Phys.\ Rev.\  D {\bf 21}, 446 (1980).
\bibitem{schnitzer}
  H.~J.~Schnitzer,
  { Confinement / deconfinement transition of large N gauge theories with  N(f)
  fundamentals: N(f)/N finite,}
  Nucl.\ Phys.\  B {\bf 695}, 267 (2004)
  [arXiv:hep-th/0402219].
\bibitem{Bas}
  P.~Basu and A.~Mukherjee,
  ``Dissolved deconfinement: Phase Structure of large N gauge theories with
  fundamental matter,''
  Phys.\ Rev.\  D {\bf 78}, 045012 (2008)
  [arXiv:0803.1880 [hep-th]].
\bibitem{Han}
  S.~Hands, T.~J.~Hollowood and J.~C.~Myers,
  ``QCD with Chemical Potential in a Small Hyperspherical Box,''
  arXiv:1003.5813 [Unknown].
  \bibitem{wadia}
  S.~R.~Wadia,
  { N = Infinity Phase Transition In A Class Of Exactly Soluble Model Lattice
  Gauge Theories,}
  Phys.\ Lett.\  B {\bf 93}, 403 (1980).
\bibitem{Hik}
  Y.~Hikida,
  ``Phase transitions of large N orbifold gauge theories,''
  JHEP {\bf 0612}, 042 (2006)
  [arXiv:hep-th/0610119].
\bibitem{Karc}
  A.~Karch and A.~O'Bannon,
  ``Chiral transition of N = 4 super Yang-Mills with flavor on a 3-sphere,''
  Phys.\ Rev.\  D {\bf 74}, 085033 (2006)
  [arXiv:hep-th/0605120].

\bibitem{Mate}
  D.~Mateos, R.~C.~Myers and R.~M.~Thomson,
  ``Thermodynamics of the brane,''
  JHEP {\bf 0705}, 067 (2007)
  [arXiv:hep-th/0701132].
\bibitem{Bas2}
  P.~Basu and S.~R.~Wadia,
  ``R-charged AdS(5) black holes and large N unitary matrix models,''
  Phys.\ Rev.\  D {\bf 73}, 045022 (2006)
  [arXiv:hep-th/0506203].
\bibitem{Yam}
  D.~Yamada and L.~G.~Yaffe,
  ``Phase diagram of N = 4 super-Yang-Mills theory with R-symmetry chemical
  potentials,''
  JHEP {\bf 0609}, 027 (2006)
  [arXiv:hep-th/0602074].
\bibitem{Mur}
  K.~Murata, T.~Nishioka, N.~Tanahashi and H.~Yumisaki,
  ``Phase Transitions of Charged Kerr-AdS Black Holes from Large-N Gauge
  Theories,''
  Prog.\ Theor.\ Phys.\  {\bf 120}, 473 (2008)
  [arXiv:0806.2314 [hep-th]].


\bibitem{Alv}
  L.~Alvarez-Gaume and S.~F.~Hassan,
  ``Introduction to S-duality in N = 2 supersymmetric gauge theories: A
  pedagogical review of the work of Seiberg and Witten,''
  Fortsch.\ Phys.\  {\bf 45}, 159 (1997)
  [arXiv:hep-th/9701069].
\bibitem{Uns}
  M.~Unsal,
  ``Phases of N(c) = infinity QCD-like gauge theories on S**3 x S**1 and
  nonperturbative orbifold-orientifold equivalences,''
  Phys.\ Rev.\  D {\bf 76}, 025015 (2007)
  [arXiv:hep-th/0703025].


\bibitem{McL}
  L.~McLerran and R.~D.~Pisarski,
  ``Phases of Cold, Dense Quarks at Large $N_c$,''
  Nucl.\ Phys.\  A {\bf 796}, 83 (2007)
  [arXiv:0706.2191 [hep-ph]].
\bibitem{Hid}
  Y.~Hidaka and R.~D.~Pisarski,
  ``Suppression of the Shear Viscosity in a ''semi'' Quark Gluon Plasma,''
  Phys.\ Rev.\  D {\bf 78}, 071501 (2008)
  [arXiv:0803.0453 [hep-ph]].






\end{thebibliography}
\end{document}